\documentclass[12pt,a4paper,oneside,headings=small]{scrartcl}
\usepackage[utf8]{inputenc}
\usepackage[english]{babel}
\usepackage{graphicx,geometry}
\usepackage{fancyhdr,mathptmx,float}
\usepackage[T1]{fontenc}

\usepackage{helvet,array}
\newcolumntype{L}[1]{>{\raggedright\let\newline\\\arraybackslash\hspace{0pt}}m{#1}}
\usepackage[x11names]{xcolor}
\usepackage{booktabs}
\usepackage{parskip}
\usepackage[format=plain,
labelfont=it,
textfont=it]{caption}

\usepackage{tikz}
\usepackage{url}
\usepackage{upgreek}
\usetikzlibrary{shapes.geometric}
\usepackage{subcaption}
\usepackage{graphicx} 
\usepackage{changes}  
\graphicspath{{figure/}}
\usepackage[comma,authoryear]{natbib}
\fancypagestyle{firstPage}{
\fancyhead[L]{}
\fancyfoot[L]{{\textit{\fontsize{9}{7}\selectfont 14$^{th}$ International Symposium on Hazards, Prevention and Mitigation of Industrial Explosions\\
Braunschweig, GERMANY - July 11-15, 2022}}}
\fancyfoot[R]{\vspace{-5mm}\includegraphics[trim=2mm 0mm 2mm 0mm,clip=true,width=0.09\textwidth]{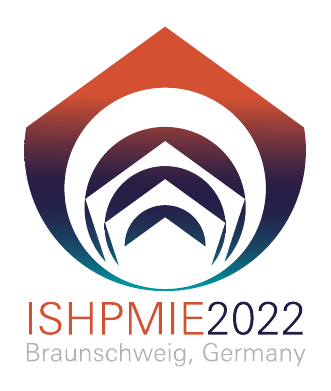}}

}

\setkomafont{section}{\normalfont\bfseries}
\RedeclareSectionCommand[
  beforeskip=-1\baselineskip,
  afterskip=.1\baselineskip]{section}
\setkomafont{subsection}{\normalfont\itshape}
\RedeclareSectionCommand[
  beforeskip=-1\baselineskip,
  afterskip=.1\baselineskip]{section}
\setkomafont{subsubsection}{\normalfont\itshape}
\RedeclareSectionCommand[
  beforeskip=-1\baselineskip,
  afterskip=.1\baselineskip]{section}

\addto\captionsenglish{
        
        }

\begin{document}
\thispagestyle{firstPage}
\begin{center}
\textbf{\Large Experimental study of humidity influence on triboelectric charging of particle-laden duct flows}\\[15pt]
Wenchao Xu, Holger Grosshans \\[5pt]
Physikalisch-Technische Bundesanstalt (PTB), Braunschweig, Germany\\
E-mail: {\color{blue} \textit{wenchao.xu@ptb.de}}\\[5pt]
\end{center}
\section*{Abstract}
Electrostatic charge on powders arises during pneumatic transport due to particle-particle and particle-surface interactions via triboelectrification. 
This is a potential threat to the safety of industrial productions and the source of numerous fires and dust explosions in the past.   
Triboelectric charges are affected by environmental conditions, such as ambient temperature and relative humidity. 
In this work, we experimentally investigated the influence of ambient humidity on the particle charge of gas-solid flows in a square-shaped duct. 
Monodisperse PMMA particles are fed into a fully developed airflow in a PMMA duct and then pass through a metallic duct section. 
The charge of particles is measured at the outlet of the metallic duct via a Faraday cup.  
By measuring the electrostatic charge under various environmental conditions, we observed that the electrostatic charge first increases with the humidity and then decreases when the humidity becomes higher. 

\medskip
\noindent
\textbf{Keywords:} \textit{electrostatics, particle charge, triboelectricity, particle-laden flow}

\section{Introduction}
Dust explosion is one of the most serious and widespread explosion hazards in the processing industry. In an industrial production involving pneumatic conveyance of powders, electrostatics require to be paid particular attention \citep{abbasi2007dust}. In the past, electrostatics has caused numerous fires and dust explosions \citep{eckhoff2003dust}.

During pneumatic transport, powders undergo undesired charging via triboelectrification due to particle-particle and particle-surface interactions. The charge can accumulate on non-conductive materials. When such an insulator with a high concentration of charge moves close to a blunt grounded conductor, a "brush discharge" can happen and ignite dust clouds \citep{larsen2001ignition}. The charged particles can also agglomerate and deposit at the inner surface of pipelines or ducts due to particle-wall adhesion. These dust deposits can result in the plugging of conveyors and lead to a frequent cleaning of the system \citep{sippola2018experimental}. Moreover, the dust deposits can accumulate heat, develop high temperatures in a spot and even cause internal smouldering fires, which serve as another ignition source for dust explosions \citep{eckhoff2003dust}.

Although the debate on the mechanism of triboelectrification is still ongoing \citep{lacks2019long}, it is widely acknowledged that the triboelectric effect is an extremely sensitive phenomenon and differs from numerous conditions, including the material properties, methods by which the materials come into contact, temperature and humidity of the surrounding environment, and other properties.

The influence of environmental conditions on triboelectric charging of polymers, i.e. ambient temperature and humidity, has been reported in many publications. 
\cite{kolehmainen2017effect} investigated the triboelectric charge of polyethylene (PE) particles in a glass container subjected to vertical vibration at different humidity levels. Applying the effective work function theory, they established a model predicting the effective work function difference regarding the humidity and showed the charge in the vibrating vessel, as well as fluidized bed, decreases non-linearly when the relative humidity increases. 
\cite{jantac2019experimental} reported that saturation charge of PE particles reduces with increasing air humidity in a shaking apparatus for a relative humidity ranging from 46\% to 67\% at 22 $^\circ \mathrm{C}$. 
Although electron pair interaction is widely acknowledged as the fundamental mechanism of triboelectric charging of polymers, \cite{nemeth2003polymer} proposed that the water adsorption of polymer particles influences the charging mechanism by introducing an additional ion conductivity into the process.
Recent research reveals that the ambient humidity can even change the polarity of the tribocharge of non-polar polymers (PC and PVC) after rubbing with Aluminum samples \citep{tilmatine2022effect}.

With the increase of humidity, the saturation charge of a particle can rapidly drop off after reaching a certain level of humidity. The experimental study by \cite{cruise2022effect} shows that the cut-off humidity level is dependent on particle size. Smaller particles are more sensitive to humidity and begin to discharge at a lower humidity level than larger particles.

Despite the plentiful studies related to the influence of ambient humidity on triboelectric charging, most of them focus on the saturation charge under various humidity. The particles in this case are charged to saturation via sufficient frictions or collisions. However, the conclusions from these studies can not be directly applied to the scenario, in which particles are charged via a limited number of impacts instead of adequate collisions.

In this paper, we present an experimental study of the influence of humidity on triboelectric charging of particles in gas-solid flows in a short square-shaped duct. In this pneumatic conveying setup, the particle samples are charged by a limited number of collisions.

\section{Experimental setup}
\subsection{Pneumatic conveying test bench}
\begin{figure}[b]
	\centering
	\begin{subfigure}[t]{0.7\textwidth}
	\resizebox{1\textwidth}{!}{
	    \begin{tikzpicture}[thick]
		\draw [->,>=latex,ultra thick] (0,0.25) node[right,above]{Air from blower} -- (1.5,.25);
		\draw [] (1.5,.0) rectangle (9,.5) node[midway] {Test duct};
		\draw [fill=gray] (7.5,.0) rectangle (9,.5) ;
		\draw [ultra thick] (7.5,0) -- (7.5,.5);
		\draw [->,>=latex,thin] (6,1.5) node[above left,align=right]{PMMA} -- (6.5,.5);
		\draw [->,>=latex,thin] (7.25,1.5) node[above]{Obstacle} -- (7.5,.5);
		\draw [->,>=latex,thin] (8.5,1.5) node[above right,align=left]{Steel} -- (8.0,.5);
		\draw [->,>=latex,ultra thick] (2.5,1.25) node[right,above]{Powder feeding} -- (2.5,.5);
		\draw [fill=gray!40,ultra thick] (9,-.5) rectangle (11,1) node[midway] {Faraday};
		\draw [thin] (11,.75) -- (11.5,.75);
		\draw [] (11.5,0.5) rectangle (14.5,2) node[align=center,midway] {Charge\\amplifier\\\& electrometer};
		\draw [->,>=latex,thin] (1.5,-.35) node[below,align=right] {300 mm} -- (2.5,-.35);
		\draw [thin] (2.5,-.2) -- (2.5,-.5);
		\draw [<->,>=latex,thin] (2.5,-.35) -- (7.5,-.35) node[below,midway,align=right] {1500 mm};
		\draw [thin] (7.5,-.2) -- (7.5,-.5);
		\draw [<->,>=latex,thin] (7.5,-.35) -- (9,-.35) node[below,midway,align=right] {500 mm};
	    \end{tikzpicture}
	}
	\caption{}
	\label{fig:rig}
	\end{subfigure}
	\begin{subfigure}[t]{0.25\textwidth}
	\resizebox{1\textwidth}{!}{
	    \begin{tikzpicture}[thick,baseline={(0,-1.2)}]
		\pgfmathsetmacro{\cubexa}{1.3}
        \pgfmathsetmacro{\cubeya}{1}
        \pgfmathsetmacro{\cubeza}{1}
        \draw[] (0,0,0) -- ++(-\cubexa,0,0) -- ++(0,-\cubeya,0) -- ++(\cubexa,0,0) -- cycle;
        \draw[] (0,0,0) -- ++(0,0,-\cubeza) -- ++(0,-\cubeya,0) -- ++(0,0,\cubeza) -- cycle;
        \draw[] (0,0,0) -- ++(-\cubexa,0,0) -- ++(0,0,-\cubeza) -- ++(\cubexa,0,0) -- cycle;
        \draw[dashed] (-\cubexa,-\cubeya,-\cubeza) -- ++(0,\cubeya,0);
        \draw[dashed] (-\cubexa,-\cubeya,-\cubeza) -- ++(0,0,\cubeza);
        \pgfmathsetmacro{\cubexb}{1.3}
        \pgfmathsetmacro{\cubeyb}{1}
        \pgfmathsetmacro{\cubezb}{1}
        \draw[fill=gray] (\cubexa,0,0) -- ++(-\cubexb,0,0) -- ++(0,-\cubeyb,0) -- ++(\cubexb,0,0) -- cycle;
        \draw[fill=gray] (\cubexa,0,0) -- ++(0,0,-\cubezb) -- ++(0,-\cubeyb,0) -- ++(0,0,\cubezb) -- cycle;
        \draw[fill=gray] (\cubexa,0,0) -- ++(-\cubexb,0,0) -- ++(0,0,-\cubezb) -- ++(\cubexb,0,0) -- cycle;
        \draw[dashed] (-\cubexa,-\cubeya,-\cubeza) -- ++(\cubexa,0,0);
        \draw[dashed] (\cubexa-\cubexb,-\cubeyb,-\cubezb) -- ++(\cubexb,0,0);
        \draw[dashed] (\cubexa-\cubexb,-\cubeyb,-\cubezb) -- ++(0,\cubeyb,0);
        \draw[dashed] (\cubexa-\cubexb,-\cubeyb,-\cubezb) -- ++(0,0,\cubezb);
        \node[cylinder, draw, shape aspect=.5, 
              cylinder uses custom fill, cylinder end fill=green!50, 
              minimum height=1cm,
              cylinder body fill=green!25, opacity=0.5, 
              scale=1.13, rotate=90](c) at (0.3,-0.35){};
        \draw [->,>=latex,ultra thick] (-2,-0.3) node[right,above]{\small Airflow} -- ++(1.5,0);
	    \end{tikzpicture}
	}
	\caption{}
	\label{fig:obstacle}
	\end{subfigure}
	\caption[]{(a) Schematic sketch of the pneumatic conveying system~\citep{grosshans2021}. The Faraday measures the charge of the powder collected by a filter at the outlet of the metallic duct. (b) The green cylinder in the enlarged sketch shows the position of the obstacle in the metallic section.}
	\label{fig:sketch}
\end{figure}
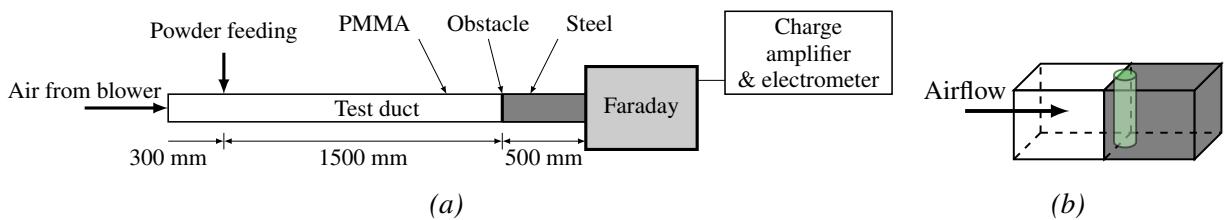

Figure \ref{fig:rig} illustrates a schematic sketch of the experiment facility. 
The present configuration consists of an air blower, a powder feeder, a square test duct, and a Faraday cage.
The air blower (Moro MHR 452) is placed at the inlet of the test duct to generate the conveying airflow. The blower is equipped with a frequency converter (Danfoss FC 51) to control the rotation speed of the blower, as well as the airflow velocity. 
A Pitot tube anemometer is installed along the channel center upstream of the particle inlet to measure the stream-wise air velocity.

The test duct includes two sections with the same inner side length of 45~mm$\times$45~mm but different materials. The upstream section is made of polymethyl methacrylate (PMMA) and has a length of 1.8~m, whereas the downstream section is made of S195T steel section with a length of 0.5~m. Particles are fed into the duct in the PMMA section at the feeding position noted in fig.~\ref{fig:rig} with a distance of 0.3~m from the airflow inlet. In addition, a steel cylindrical obstacle with a diameter of 17~mm is placed at the middle of the entrance of the metallic section (see fig.~\ref{fig:obstacle}). The obstacle functions as a generic representation of components that are frequently built in pneumatic conveying systems, such as sensors, screws, or other pipes. The obstacle together with the metallic duct is grounded so that the electric potential of the metal surface remains constant. 

\subsection{Measurement equipment}
To measure the charge of the particles, a 50 $\mathrm{\upmu m}$ pore size filter bag is attached at the end of the duct and meanwhile covered by a Faraday cup (Monroe Electronics 284/22A). The Faraday cup is connected to a charge amplifier (PCB 44302). Particles are separated from the airflow by the filter and their charge is measured by the charge amplifier with a systematic error of 2\%. It is worth mentioning that the charge of the electrical circuit leaks consistently. During the experiment, the time-scale of the charging measurement was set to be significantly smaller than the charge leakage time-scale of the electrical circuit, so that the error due to circuit leakage was negligible.

To reduce the error introduced by the leakage, the time-scale of the charging measurement is required to be significantly smaller than the charge leakage time-scale of the electrical circuit.

The temperature and relative humidity are measured by a high accuracy temperature and humidity sensor (Lascar EL-WiFi-TH+) with a tolerance of $\mathrm{\pm 0.2~^\circ C}$ for temperature and $\mathrm{0.1\%}$ for relative humidity.

\subsection{Experimental procedure}
In the experiment, the airflow velocity is regulated to $13~\mathrm{m/s}$. Due to the character of the blower, the velocity fluctuates persistently with $\pm0.15~\mathrm{m/s}$ during the experiment. This velocity is measured by the Pitot tube anemometer at the centre-line, which corresponds to the maximum velocity in the cross-section of the airflow. The mean flow velocity, i.e. the mean conveying velocity, can be estimated by multiplying a coefficient of 0.8 \citep{susanti2020measurement}. 

After the airflow is stabilized, particle samples are fed into the duct. In this study, we use monodisperse spherical PMMA particles from EPRUI Biotech with a diameter of 100 $\mathrm{\upmu m}$. The particles have a material density of $\mathrm{1.15~g/cm^3}$ and a bulk density of $0.8~\mathrm{g/cm^3}$.
To avoid measurement error due to charge leakage and meanwhile not exceed the measurement range of the charge amplifier, we inject $\mathrm{0.5~g-1.5~g}$ particles into the PMMA duct within less than 10 s in each measurement.

Considering the short duration of each measurement, the feeding of particles is accomplished manually with a syringe instead of a vibrating feeder. Before and after each injection, the weight of the syringe is measured by a precise scale (Kern PCB 3500-2) with $0.01~\mathrm{g}$ accuracy and the differential of the two measurements returns the weight of the fed powder. After a set of measurements, the particles collected by the filter are weighted to verify the measurement during each injection. Because the conveying system uses a positive pressure configuration, part of the particles might be blown away during the feeding process. The fine particles may also agglomerate on the inner wall of the duct. In our experiment, the ratio between the collected mass and the fed mass is higher than $95\%$, i.e., the mass loss during the feeding and the pneumatic conveying is less than $5\%$. Because the majority of the mass loss occurs during the feeding process due to the positive pressure in the conveying system, the influence of the mass loss on the charge measurement should be small.

\section{Experimental repeatability}
Being an extremely sensitive process, triboelectric charging is influenced by various conditions and parameters. To ensure the repeatability of the experiment, the whole process needs to be precisely controlled. In this section, we discuss the parameters we suspect to affect the repeatability of the charge measurement the most, namely the conveying velocity, powder mass flow rate, and the state of the duct's surface.

\subsection{Conveying velocity}
\label{sec:convey_velocity}
The impact velocity of particles on the conveying system, which increases with the conveying velocity, affects the electrostatic charging. 
This parameter is strongly correlated to the practical production process, since a high conveying speed during pneumatically transporting powdery materials is usually preferred to avoid deposition and stagnation of powders. 

\begin{figure}[tb]
	\centering
    \includegraphics[trim=0mm 0mm 0mm 0mm,clip=true,width=0.5\textwidth]{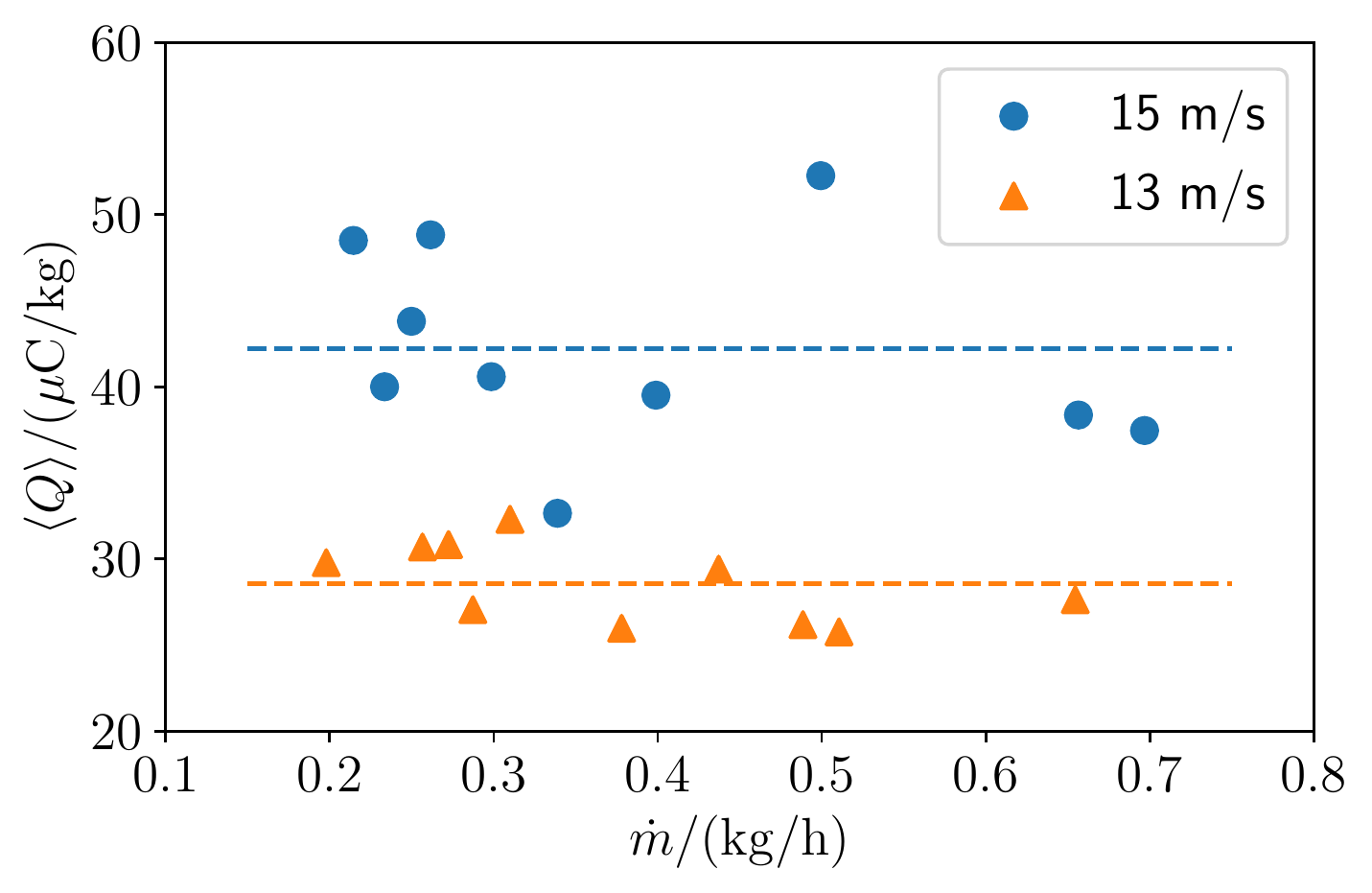}
	\caption{Particle charge with different airflow velocities at the same ambient condition ($RH= 54\%$, $T= 16.3~^\circ C$). The dashed line represents the average charge at each velocity. }
	\label{fig:velocity}
\end{figure}

Figure \ref{fig:velocity} depicts the effect of conveying velocity on particle charge under the same environmental condition. 
For an air velocity of 15~m/s, the particles obviously receive more charge than the lower velocity of 13~m/s. 
This result is consistent with the experiments reported by \cite{matsusaka2000electrification} and \cite{watanabe2006measurement}. 
Higher conveying velocity contributes to an increase of the normal component of the impact velocity, which, according to Hertz theory, proportionally increases the maximal contact area of an elastically deformed particle impacting against a plane \citep{watanabe2006measurement}. Consequently, it increases the contact area between the particle and the  and promotes triboelectrification.

Therefore, to receive repeatable results, the conveying velocity should remain constant during all the measurements.

\subsection{Powder mass flow rate}
To ensure a small feeding quantity and a short feeding duration in each measurement, we injected the powders into the duct manually using an syringe. In this case,the mass flow rate can not be precisely controlled and inevitably varies upon each attempt. Our previous study reported that the mass flow rate affects the powder charging in the particle-laden airflow \citep{grosshans2021}. According to the numerical simulations presented by \cite{grosshans2021}, the wall-normal velocity of particles is faster for a higher mass flow rate in the dilute particle-laden flow. As a consequence, triboelectric charging is also promoted, which is analogous to the procedure of increasing conveying velocity. Moreover, when the mass flow rate further increases, a high concentration of charged particles can give rise to the surrounding electric field, and, according to the charge relaxation theory \citep{matsuyama1995electrification}, reduce the charge a particle can hold.


In the experiment, to reduce the influence of varying mass flow rate, we repeated multiple times measurements for each experiment and meanwhile tried to cover a wide range of mass flow rates. 

\subsection{State of the duct's surface}
We suspect the state of the conveying duct's surface, cleanliness and charge spots, to have a major impact on the experimental repeatability. 

Fine particles can adhere or deposit on the inner wall of the pneumatic conveying system. With the presence of electrostatic force, particles adhere stronger \citep{hays1995adhesion}. The adherence on the surface not only increases the energy loss during pneumatic transport by increasing surface friction, but also influences the triboelectric charge. 

\begin{figure}[tb]
    \begin{subfigure}{0.46\textwidth}
         \centering
         \includegraphics[trim=0mm 0mm 0mm 0mm,clip=true,width=1\textwidth]{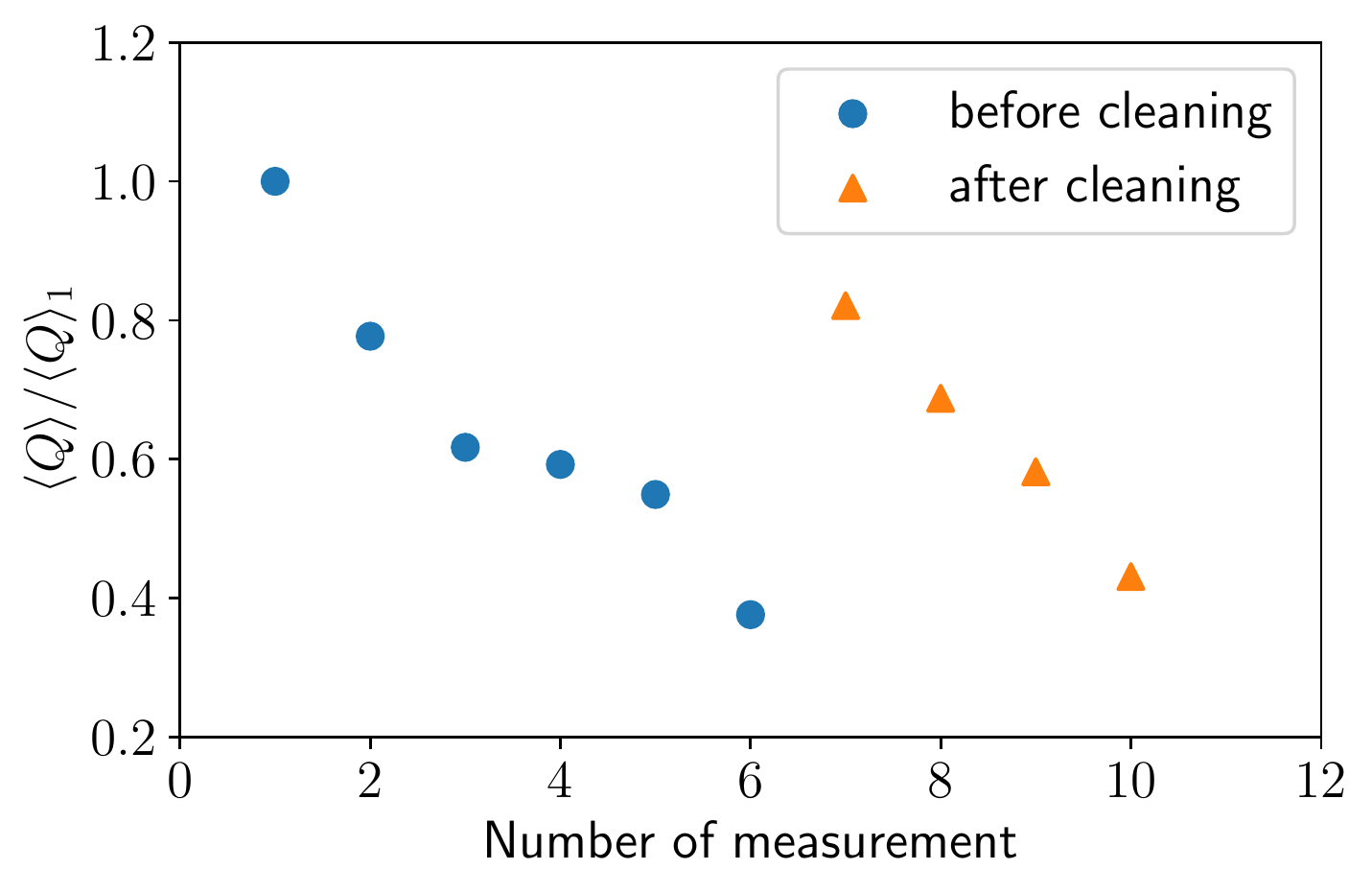}
         \caption{}
         \label{fig:clean1}
     \end{subfigure}
     \quad
     \begin{subfigure}{0.46\textwidth}
         \centering
         \includegraphics[trim=0mm 0mm 0mm 0mm,clip=true,width=1\textwidth]{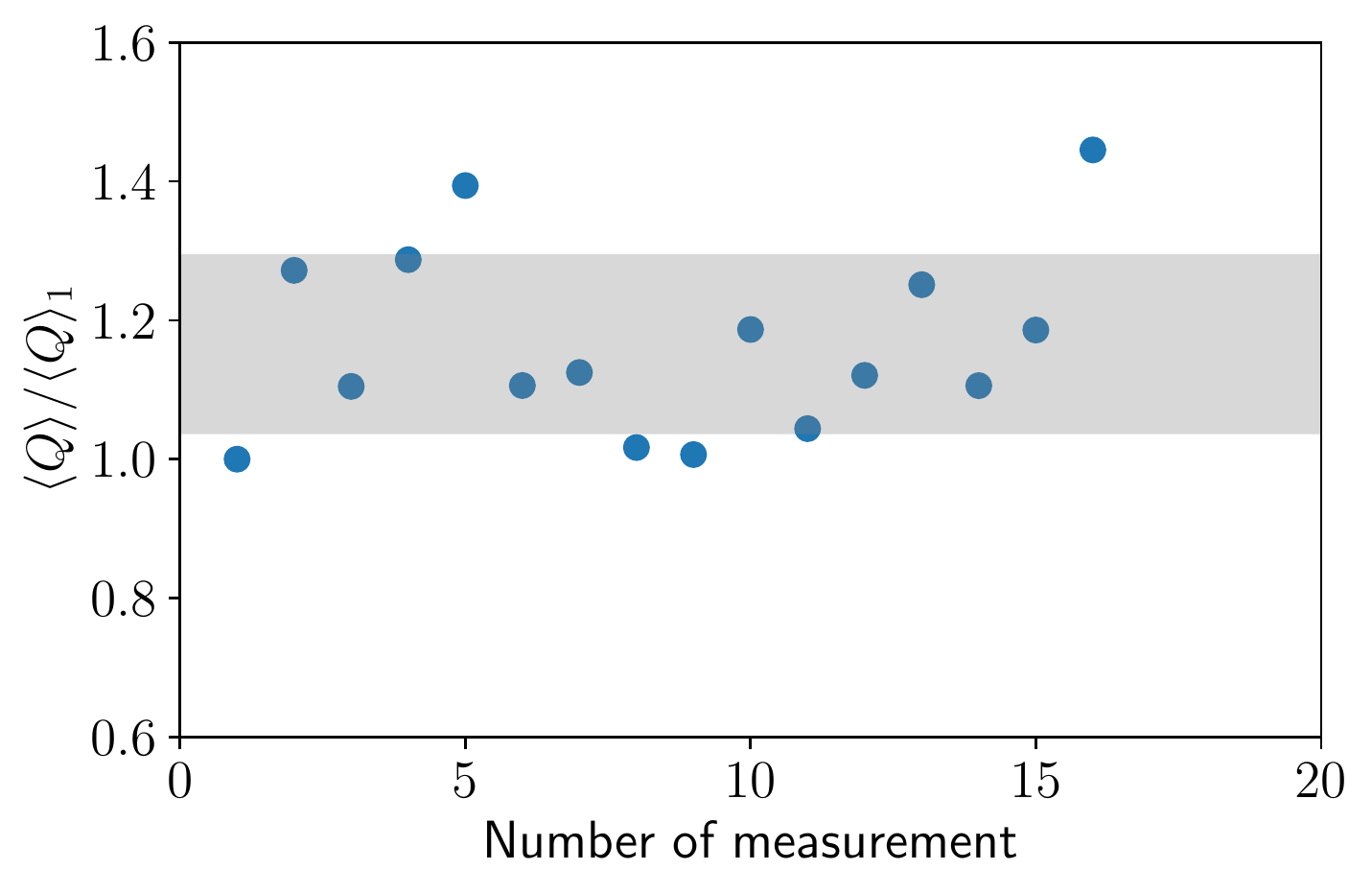}
         \caption{}
         \label{fig:clean2}
     \end{subfigure}
	\centering
	\caption{Particle charge as a function of the measurement sequence. The charge of each measurement $\left<Q\right>_i$ is normalized by the first measurement $\left<Q\right>_1$. The gray area in (b) marks the measurements within one standard deviation of the mean charge.}
	\label{fig:clean}
\end{figure}

In our experiment, we occasionally observed significant drops of particle charge during successive measurements, see blue dots in fig.~\ref{fig:clean1}. After properly cleaning the duct with ethanol and drying the test rig, the measured specific charge is recovered temporally and then continues decreasing (orange triangles in fig.~\ref{fig:clean1}). Nevertheless, such a dramatic decrease of the charge seldom occurred in the experiment compared to the "normal" experiment (see fig.~\ref{fig:clean2}), in which the measured charging is not correlated with the cumulative number of measurements.

To avoid the influence of surface contamination and deposition, the duct was cleaned with ethanol regularly. The experimental data, when such a phenomenon occurred, were regarded as invalid data and excluded from subsequent discussions. 

\section{Charging distribution in the conveying duct}
In the reported configuration, particles are released upstream in the PMMA duct section, so that the particle-laden airflow is fully developed before entering the metallic section. The volumetric flow rate of the particles ($10^{-1}~\mathrm{cm^3/s}$) is much smaller than of the air ($10^{5}~\mathrm{cm^3/s}$). In such a dilute phase gas–solid duct ﬂow, the effect of particle–particle interactions on particle charging is negligible. Each particle can freely collide with the inner wall without hindering from surrounding particles. Although the study focuses on triboelectrics in the metallic section, charging in the PMMA duct is worth to be noticed. 

\begin{figure}[tb]
	\centering
	\begin{subfigure}{0.46\textwidth}
         \centering
         \includegraphics[trim=0mm 0mm 0mm 0mm,clip=true,width=1\textwidth]{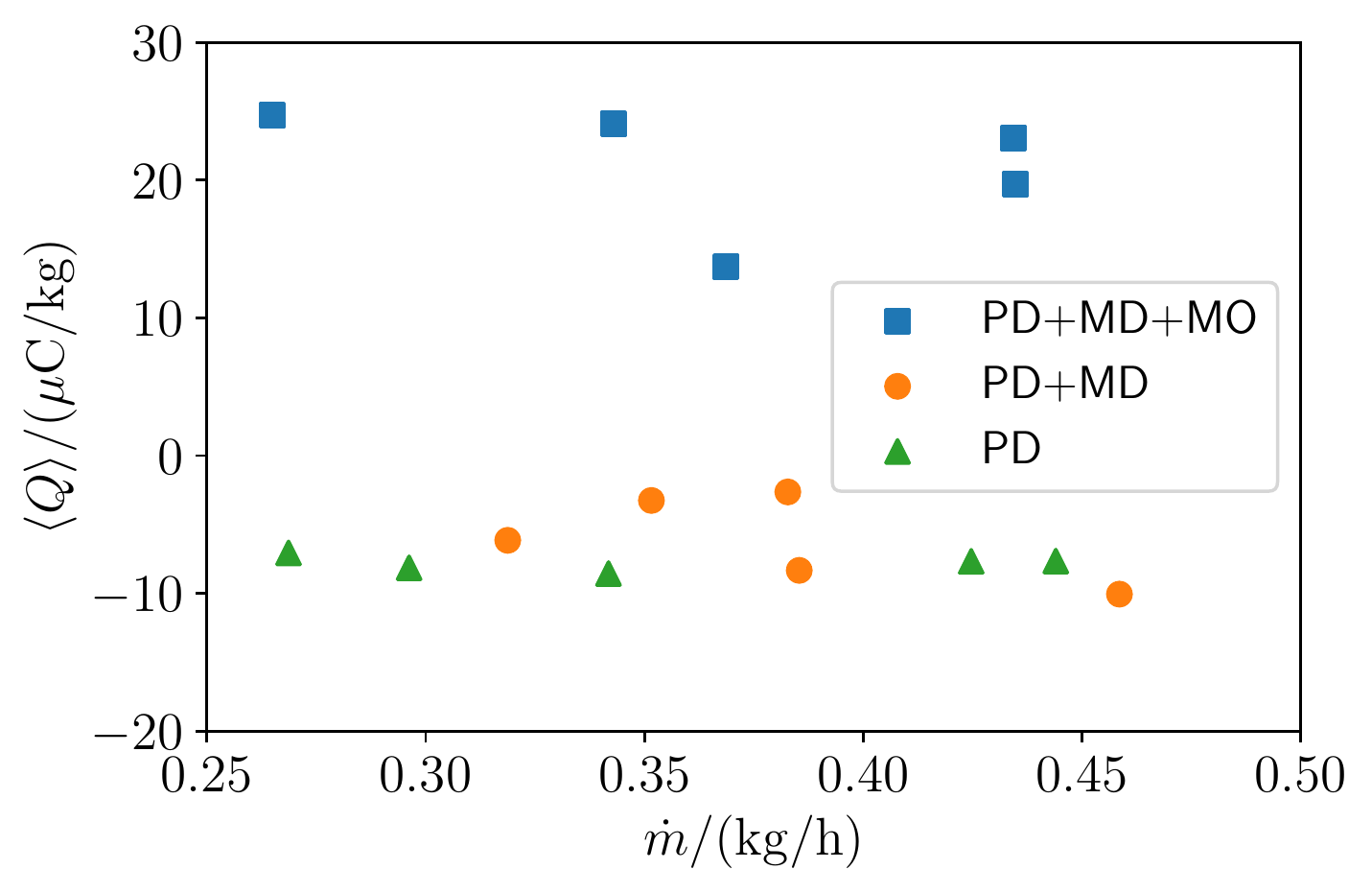}
         \caption{}
         \label{fig:cyl_exp}
     \end{subfigure}
     \quad
     \begin{subfigure}{0.46\textwidth}
         \centering
         \includegraphics[trim=0mm 0mm 0mm 0mm,clip=true,width=1\textwidth]{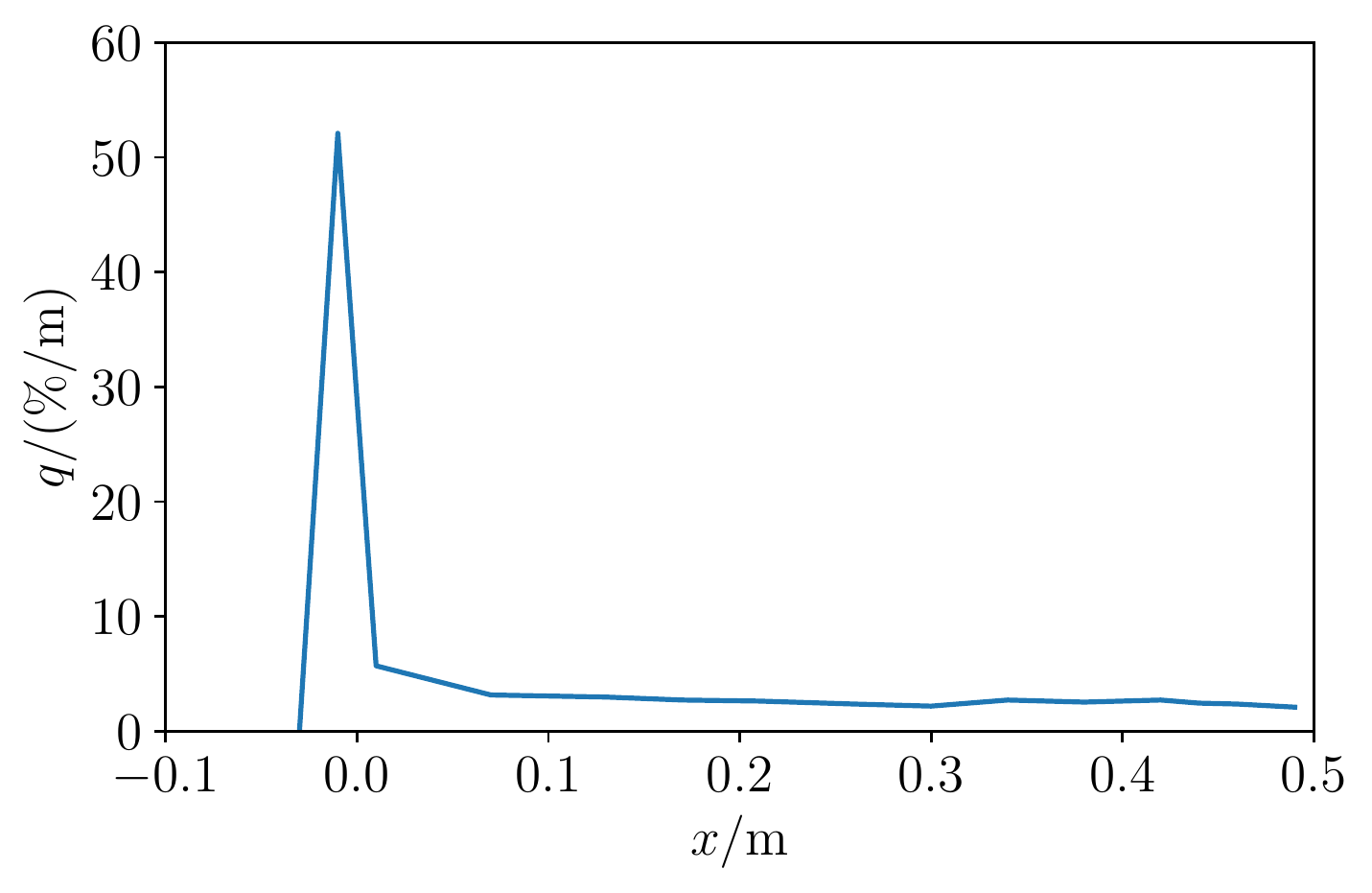}
         \caption{}
         \label{fig:cyl_sim}
     \end{subfigure}
	\caption{(a) Particle charging with different setups. Squares: normal setup with PMMA duct (PD), metallic duct (MD), and metallic obstacle (MO); circles: test duct with PMMA duct and metallic duct (no obstacle); triangles: test duct only with the PMMA section. (b) Charge generation per meter as a fraction of the total powder charge along the streamwise direction with mass flow rate 0.091 kg/h and centre-line velocity 14.7 m/s from numerical simulation.}
	\label{fig:cylinder}
\end{figure}

Figure \ref{fig:cyl_exp} shows the charge-to-mass ratio of particles, $\left< \mathrm{Q} \right>$, measured under different configurations with the same parameter. The blue squares represent the standard configuration, the orange dots correspond to the setup without the metallic obstacle at the entrance of the metallic duct, and the green triangles are the setup in which the metallic components, including the obstacle and the metallic duct, are completely removed. 
it is noticed that the charging differences between the orange dots and green triangles are very small, which indicates the triboelectrical charging between particles and the metallic duct is very weak. The blue squares, on the other hand, denote a much higher level of triboelectrical charging for the setup with the metallic obstacle compared to the setups without the obstacle.  
Obviously, the metallic obstacle plays an essential role in the triboelectrical process. 

This result is consistent with our numerical simulations. Figure \ref{fig:cyl_sim} shows the percentage of particle charges in the powder flow along the streamwise direction from simulations using our open-source CFD tool pafiX \citep{pafix}.
Pafix uses an Eulerian–Lagrangian approach, in which the fluid and the particles are solved respectively in Eulerian and Lagrangian framework. The triboelectrification is calculated with an empirical charging model, which approximates the impact charge of a particle in the simulation directly from similar impacts measured in a single-particle experiment \citep{grosshans2021}. A detailed description of the mathematical model and numerical methods implemented in pafiX was given by \cite{grosshans2021effect}. 
According to fig.~\ref{fig:cyl_sim}, more than half of the triboelectric charging occurs at contacts between the particles and the cylindrical obstacle.


This can be attributed to the following two reasons. First, due to the short length of the metallic duct (0.5 m), the particle-wall contacts in the metallic section occur less frequently on the inner wall of the duct than on the cylindrical obstacle. Second, the wall-normal velocity of particles is much higher when colliding with the cylinder compared with contacting the walls parallel to the stream direction. 

Moreover, we observed that the PMMA particles are charged negatively in measurements without the cylindrical obstacle, see fig.~\ref{fig:cyl_exp}, which suggests the PMMA particles acquire electrons when colliding with a wall made by the same material. This phenomenon is analog to the scenario where particles of different sizes and the same insulating material contact with each other. Large particles usually tend to charge positively whereas small particles charge negatively \citep{waitukaitis2014sizedependent}. The PMMA particles can be regarded as "small particles" when colliding on the PMMA duct, therefore they get a negative charge after contact. 

\section{Influence of humidity}
\begin{figure}[tb]
	\centering
	\begin{subfigure}{0.46\textwidth}
         \centering
         \includegraphics[trim=0mm 0mm 0mm 0mm,clip=true,width=1\textwidth]{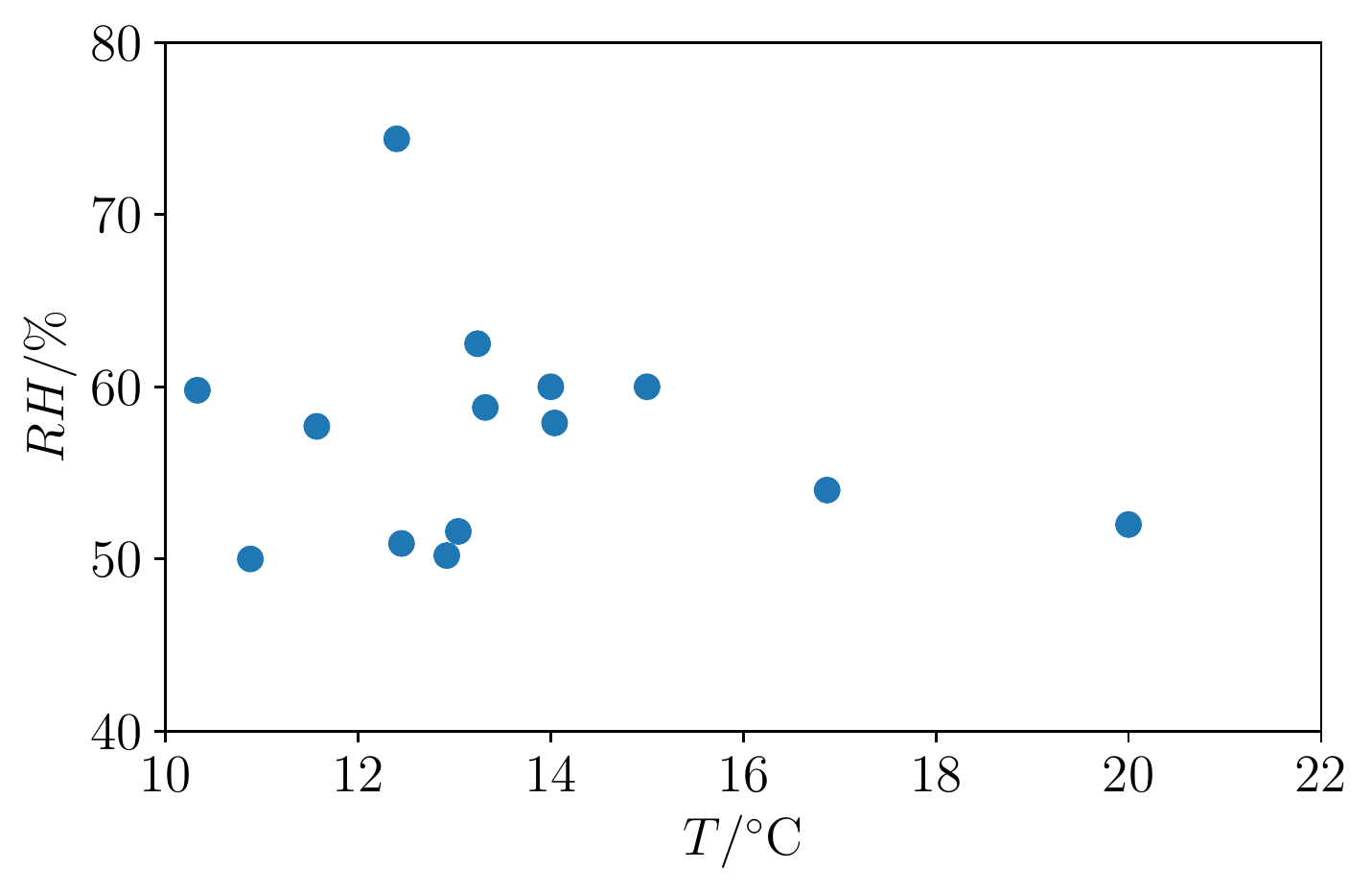}
         \caption{}
         \label{fig:rh}
     \end{subfigure}
     \quad
     \begin{subfigure}{0.46\textwidth}
         \centering
         \includegraphics[trim=0mm 0mm 0mm 0mm,clip=true,width=1\textwidth]{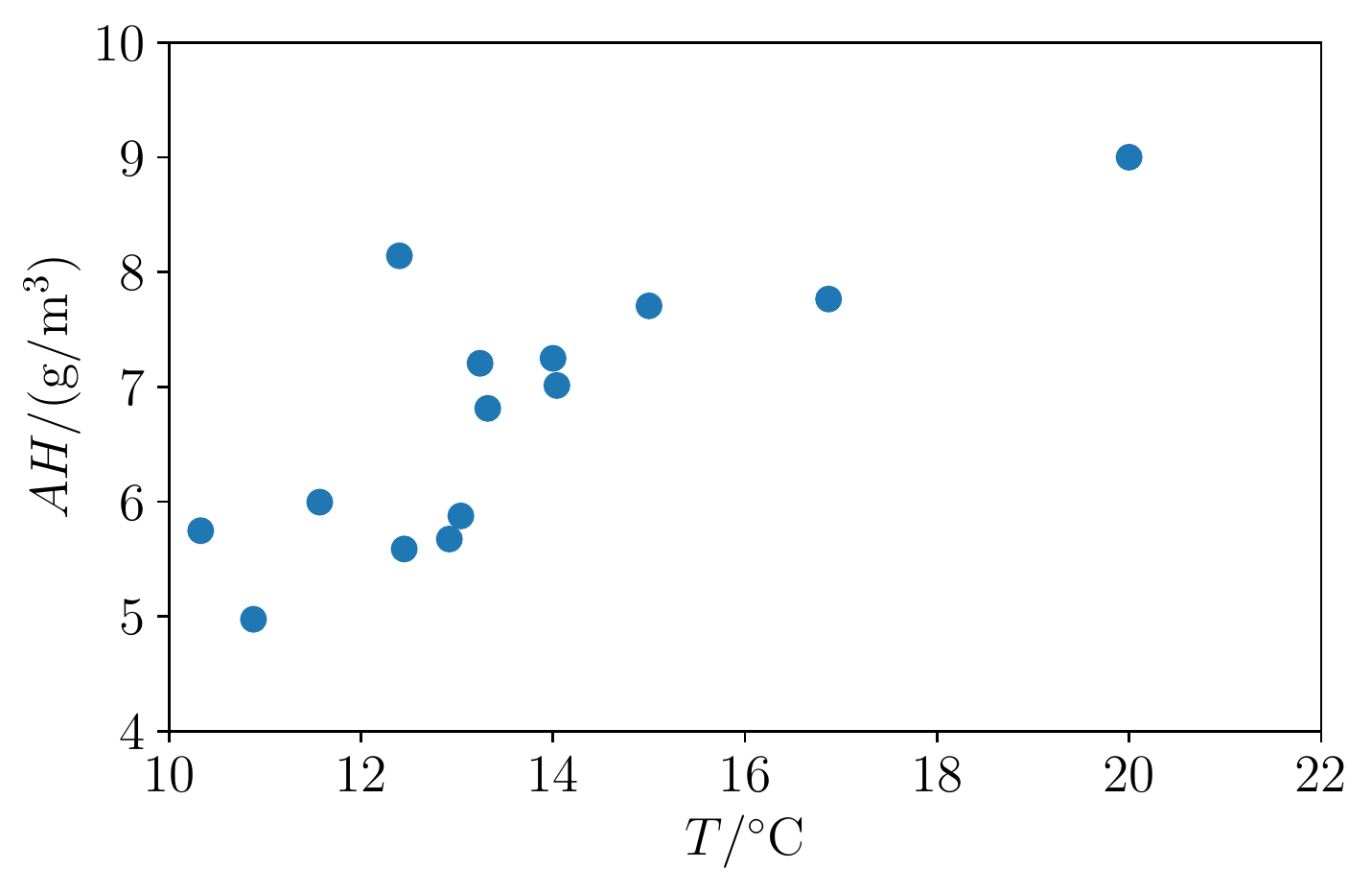}
         \caption{}
         \label{fig:ah}
     \end{subfigure}
	\caption{(a) Distribution of temperature and relative humidity of the experiment attempts; (b) distribution of temperature and absolute humidity of the experiment attempts.}
	\label{fig:humid_range}
\end{figure}

The laboratory for the experiment is not climate-controlled so that the room temperature is affected by the day-to-day weather variation. To study the influence of humidity on particle charge, we carefully recorded the temperature ($T$) and relative humidity ($RH$) variations in the lab and performed experiments under different weather conditions. Due to the slow variation of the room temperature and the short duration of each experiment (around 30 min), change of the ambient condition during an experiment can be neglected and the temperatures of the test rig and the room are considered identical. Thus, we were able to measure the inﬂuence of the air properties on the particle charge for a range of $RH$ = $50\%~ - ~74.4\%$ and $T$ = 10.3~$^\circ \mathrm{C}~-~20~^\circ \mathrm{C}$. The temperature and relative humidity during the experiments are plotted in fig.~\ref{fig:rh}.
To exclude the influence of the fluctuating room temperature on the moisture content in the ambiance, we introduce the absolute humidity (AH), which is obtained from the following empirical equation \citep{ogino2019triboelectric}:
\begin{equation}
    AH = \frac{217}{T+273.15}\times6.1078\times \exp\left( \frac{17.2694T}{T+237.3} \right)\times \frac{RH}{100}
\end{equation}
where $T$ is the room temperature in $^\circ \mathrm{C}$ and $RH$ is in $\%$. The obtained $AH$ is in the unit of $\mathrm{g/m^3}$. As shown in fig.~\ref{fig:ah}, the experiments are performed in an absolute humidity range of 4.97$~\mathrm{g/m^3}$ - 8.98$~\mathrm{g/m^3}$. 

\begin{figure}[tb]
     \centering
     \begin{subfigure}{0.46\textwidth}
         \centering
         \includegraphics[trim=0mm 0mm 0mm 0mm,clip=true,width=1\textwidth]{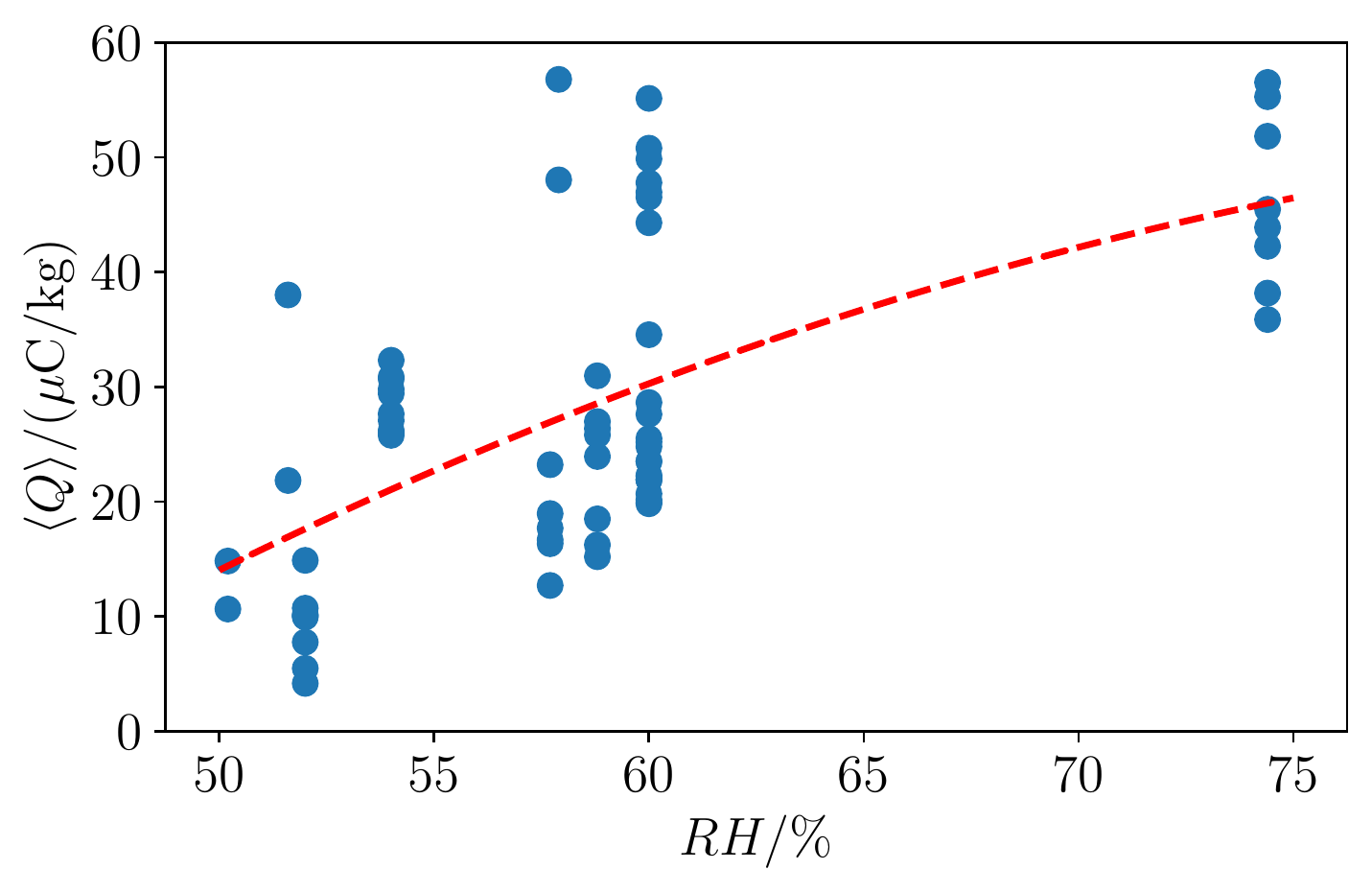}
         \caption{}
         \label{fig:charge_humidity_rh}
     \end{subfigure}
     \quad
     \begin{subfigure}{0.46\textwidth}
          \includegraphics[trim=0mm 0mm 0mm 0mm,clip=true,width=1\textwidth]{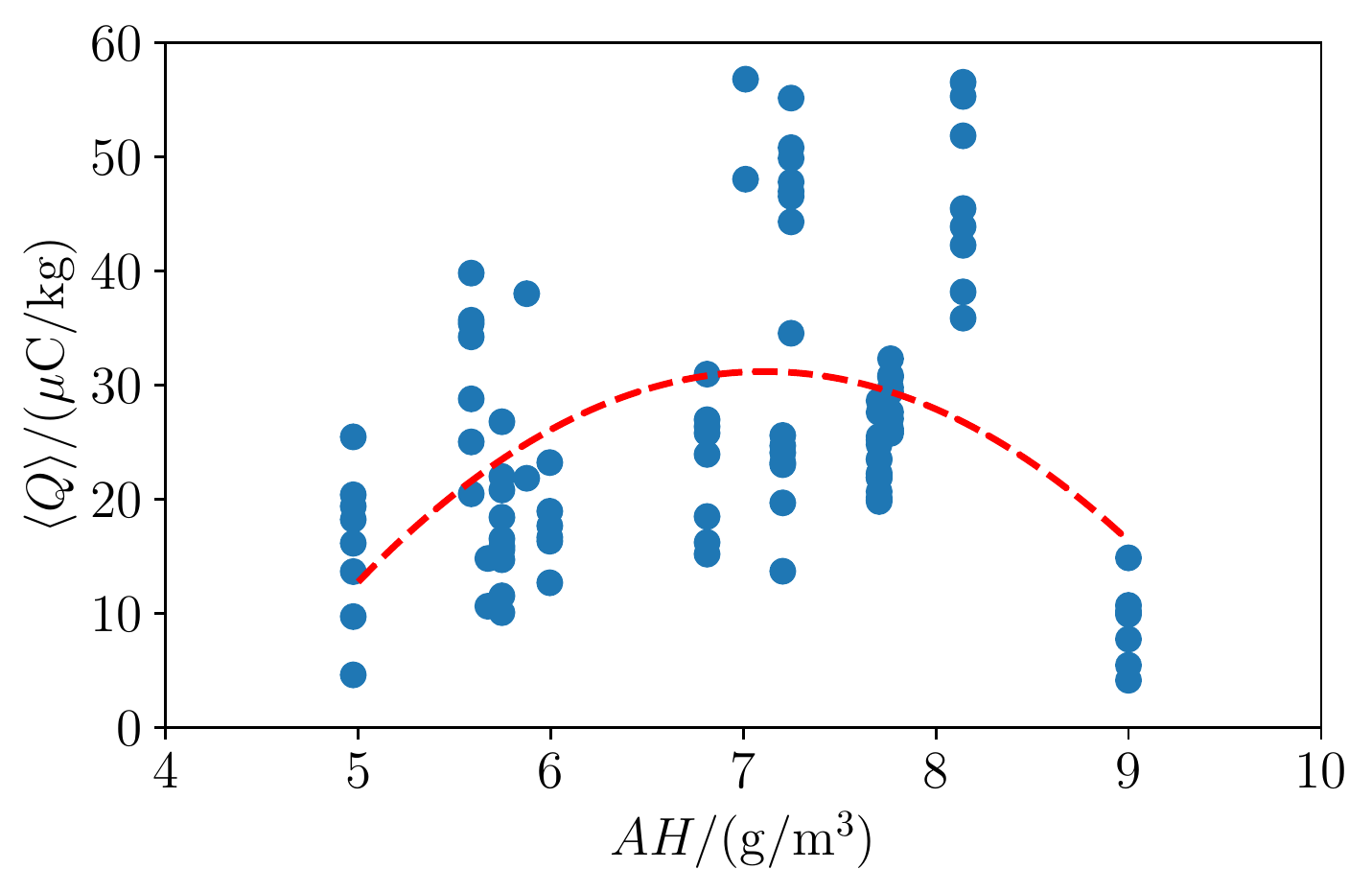}
          \caption{}
          \label{fig:charge_humidity_ah}
     \end{subfigure}
     \caption{Charge-to-mass ratio as a function of (a) relative humidity and (b) absolute humidity. Each dot represents one measurement attempt. The red dashed curve represents a second order least squares polynomial fit.}
    \label{fig:charge_humidity}
\end{figure}
The experimental results are presented in fig.~\ref{fig:charge_humidity}, which depicts charge-to-mass ratio against relative humidity and absolute humidity. In the diagram, each point represents one measurement by feeding a certain portion of particles in the duct flow. 
The feeding mass is not constant but varies with each attempt. The dots at the same humidity correspond to repeated measurements for each environmental condition.

In fig.~\ref{fig:charge_humidity}, the charge-to-mass ratio is solely described by relative humidity and shows to be increasing monotonically with the rising relative humidity. However, given the varying room temperature, the same relative humidity by different room temperatures does not represent the same moisture content in the ambient circumstance. Therefore, applying the relative humidity exclusively is not sufficient to compare the influence of the ambient moisture.

Figure \ref{fig:charge_humidity_ah} shows the charge-to-mass ratio as a function of absolute humidity, which is a measure of the actual amount of moisture in the air regardless of the air's temperature. As indicated with the polynomial fit curve, the charge-to-mass ratio first increases and then decreases with increasing absolute humidity. At first glance, the result seems to exhibit a trend inconsistent with the prevalent theory, that increased humidity reduces the saturation charge of polymer particles and increases the decay of electrostatic charge. However, the scenario in our experiment is different from the typical experiments with shakers or fluidized beds. Particles in the test duct acquire only a limited number of contacts or collides with other particles/surfaces. Therefore, the particles passing through the test rig might not be sufficiently charged and the measured charge must not be equal to the saturation charge. 

\cite{nemeth2003polymer} proposed a theory of the function of water molecules on tribocharges, that the charge transfer is dominated by electrons at low humidity, whereas adhered water and ions in water promote triboelectric charging with increased humidity. This theory well explains the relationship between the specific charge and the humidity. Electron transfer is the fundamental process of polymer charging. At lower humidity, the tribocharging is marginally influenced by water molecules in the ambient. With increased humidity, the water can swell the surface of the polymer or form absorption layers onto polymer surfaces. Due to the auto-dissociation of water and the solvation of impurities on the particle surface during production, this formed water-containing layer introduces ionic species into the charging process. The ions can decrease the surface conductivity and intensify the triboelectric charging.

The thickness of the water-absorption layer is correlated to the amount of available water in the surrounding air. When the humidity further increases, the layers get thicker, thus, decrease the upper limit of the surface charge reduces and impedes the particles from obtaining more charge. Therefore the specific charge drops at $AH = 9~\mathrm{g/m^3}$. 

Moreover, PMMA is sensitive to the moisture in the surrounding atmosphere due to its good ability in absorbing water \citep{nemeth2003polymer}. Therefore the PMMA particles are able to quickly form water-containing swollen layers by absorbing water molecules from the air. 

\begin{figure}[tb]
\centering
     \begin{subfigure}{0.45\textwidth}
         \includegraphics[width=0.9\textwidth]{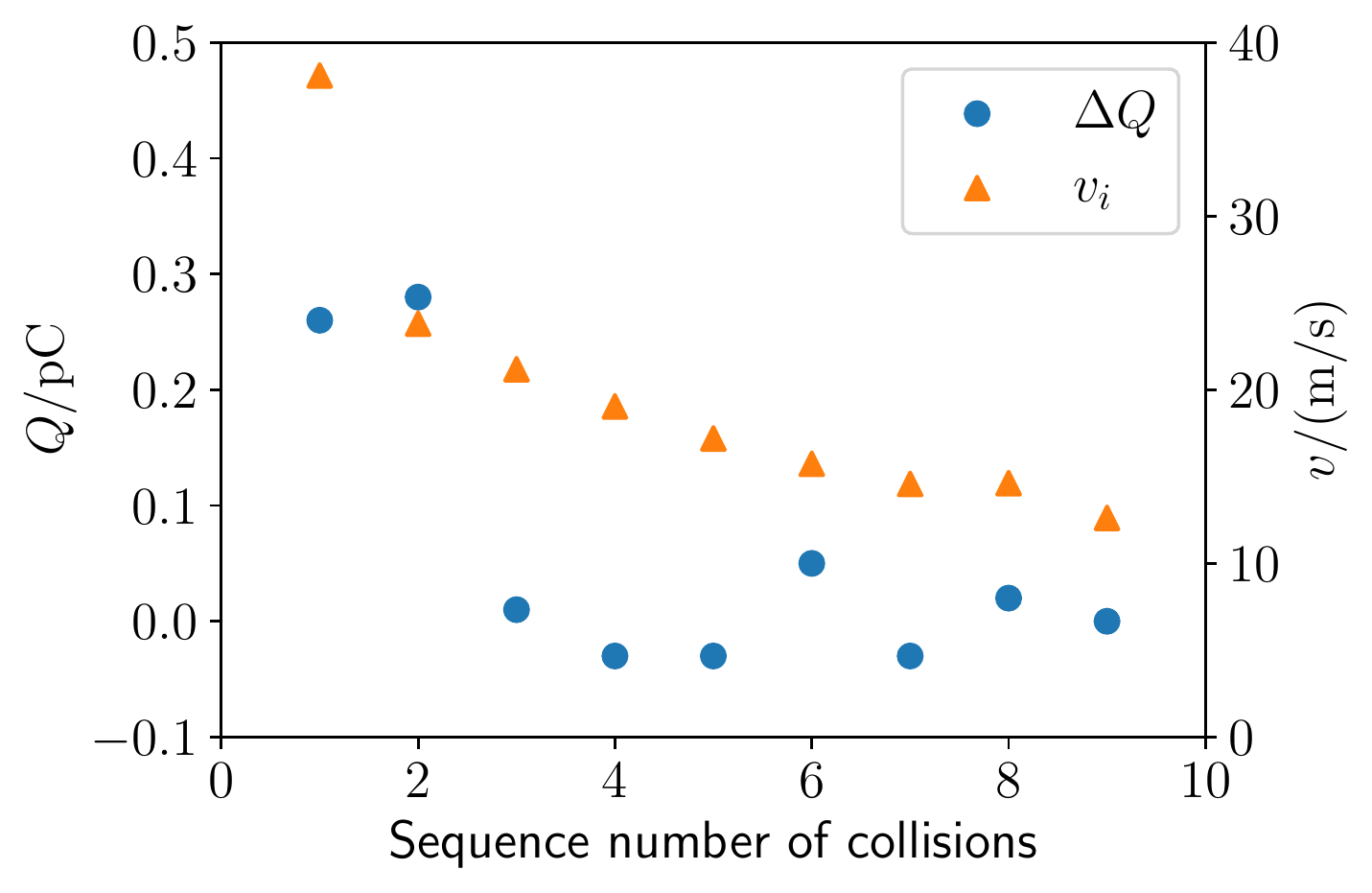}
         \caption{}
         \label{fig:single_particle}
     \end{subfigure}
     \begin{subfigure}{0.45\textwidth}
         \includegraphics[width=0.9\textwidth]{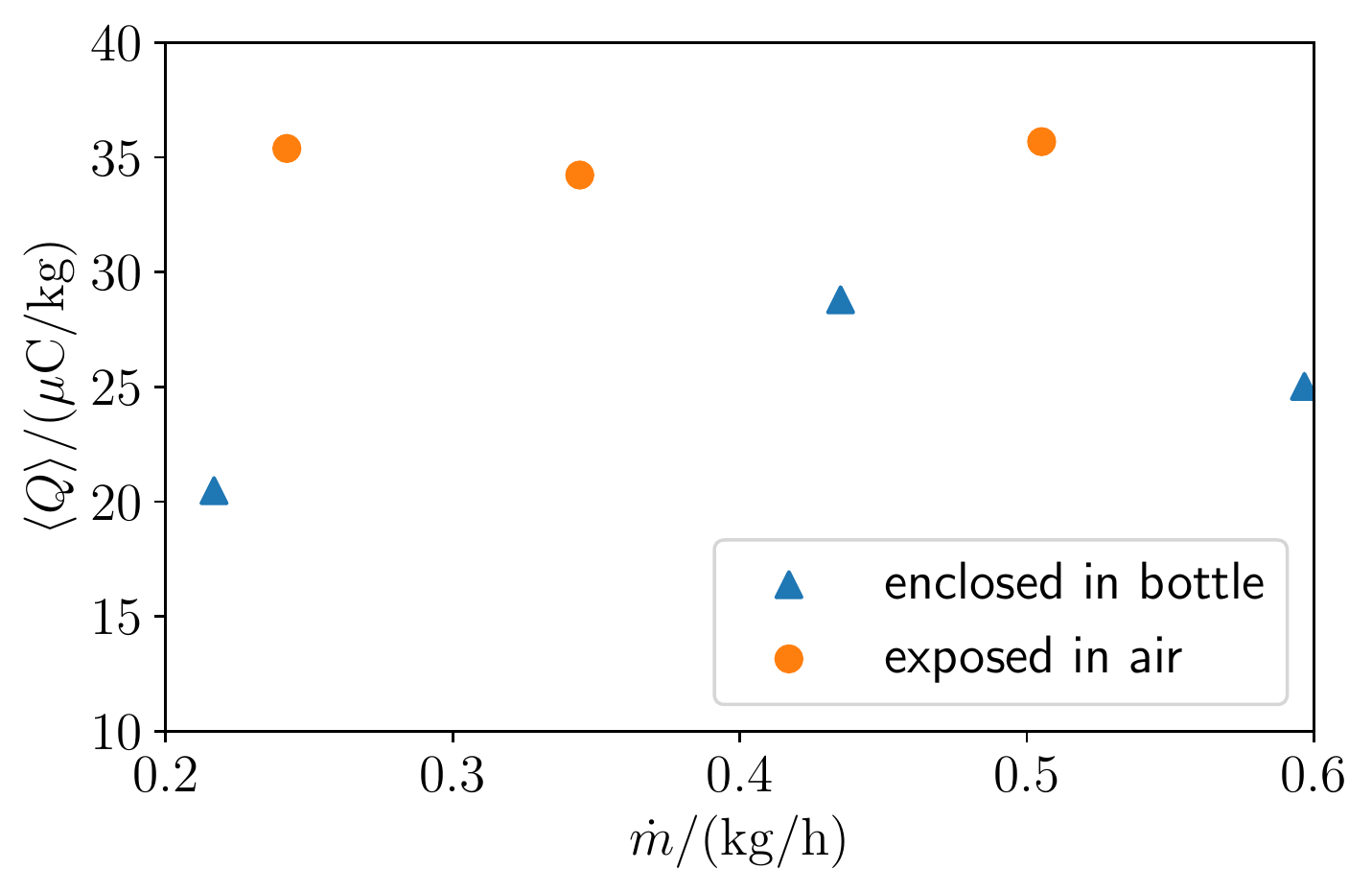}
         \caption{}
         \label{fig:particle_compare}
     \end{subfigure}\\
     \caption{(a)Charge and velocity of a 200 $\mu m$ PMMA particle in a single particle experiment (adapted from \cite{grosshans2021}). (b)Charge-to-mass ratio of particles stored in different conditions. Blue triangles: particles stored in a hermetical container; yellow dots: particles exposed to air for 24 hours. ($RH= 50.9\%$, $T= 12.5~^\circ C$). }
\end{figure}

According to our previous single-particle experiment \citep{grosshans2021}, in which a 200 $\mathrm{\upmu m}$ PMMA particle was shot onto two parallel metallic plates, the charging of the particles heavily relies on the first few contacts when the particle fast collides with a surface. Figure \ref{fig:single_particle} displays the charge and velocity of one particle in a succession of collisions. The charge transfer during the first two collisions is much stronger than the following collisions. This behavior suggests that the particle is saturated after the second impact. Such an efficient charge transfer might be also related the water-absorption layer on the particle surface, which increases the efficiency of the triboelectrical charge during an impact. 

This effect can be also elucidated using the classical condenser model. In the condenser model, the charge transferred in a sequence of impacts is $\mathrm{d} q_{c}/\mathrm{~d} n=k C V$, where $n$ is the number of impacts, $k$ is the electrification efficiency, $C$ is the capacitance between the contact bodies and $V$ is the potential difference between the contact bodies \citep{matsusaka2010triboelectric}. The existence of the water layer promotes the surface charge via increasing the electriﬁcation efficiency $k$, therefore increases the charge during a single contact.

To further validate the influence of the moisture on tribocharging, we prepared two particle samples for a comparison experiment. Both samples applied identical 100 $\mathrm{\upmu m}$ spherical PMMA particles. One sample was stored in a shallow glass tray and exposed in the air at a room temperature of $12~^\circ \mathrm{C}~-~16~^\circ \mathrm{C}$ and relative humidity around $50\%$ for 24 hours, while the other sample was stored in a dry sealed container. 

Both of the samples were fed into the test duct with the same operating parameter and the specific charge after conveying was measured, see fig.~\ref{fig:particle_compare}. As expected, the water-contained layer due to the long-time exposure in the air results in a higher particle charge compared to the particles stored in the closed container. Moreover, the charge of the previous sample is seemly more stable, which implies the particles stored in a moister environment are closer to saturation at the outlet of the duct.

\section{Conclusion and outlook}
In this work, we reported a series of experiments on triboelectric charging of PMMA particles in a pneumatic conveying duct under various temperatures and humidities. 
To investigate the influence of humidity at different temperatures, instead of relative we applied absolute humidity, which provides a quantitative indication of the moisture content in the atmosphere.

According to the experimental results, the particle charge first increases with the rise of the absolute humidity and then decreases when the humidity exceeds a threshold. This phenomenon can be attributed to the surface layer of PMMA particles formed by absorbed water molecules from the surrounding air. This water-containing layer increases surface conductivity, promotes triboelectric charging, and meanwhile decreases the saturation of surface charge. In the conveying system with a limited number of collisions, the particles are not sufficiently charged, therefore the existence of water-containing layers increases the triboelectric charge. However, when the humidity becomes higher, the saturation charge decreases dramatically, thus resulting in a lower charge.

Moreover, we discussed several effects that can hinder reproducing the experiments. Changing conveying velocity and mass flow rate can influence particle charging. Adhesion on the inner surfaces of the conveying duct may reduce the charge of particles. For a repeatable experiment, the conveying velocity and the mass flow rate should remain consistent through all measurements, meanwhile, the duct should be cleaned regularly for a consistent surface state. Our results also reveal that increasing conveying velocity can raise risks of dust explosion due to stronger triboelectrification, particularly when there are surfaces or obstacles on the flow path, on which the particles can collide with a high wall-normal velocity. On the other hand, from the perspective of preventing static electricity, direct wall-normal collisions should be avoided by designing a pneumatic system.

The experiment shows that in a system where the particles are not sufficiently charged, increasing the humidity can intensify triboelectric charging. However, this study has only applied 100 $\upmu \mathrm{m}$ PMMA particles for the measurements. It remains an open question if particles with different sizes or a material with different abilities to take up water can change the behavior of triboelectric charging in response to varying humidity. This will be part of a future investigation.

\section{Acknowledgement}
This project has received funding from the European Research Council (ERC) under the European Union’s Horizon 2020 research and innovation programme (grant agreement No. 947606 PowFEct).
{\setlength{\bibsep}{1pt}
\bibliography{bibliography}}
\end{document}